\documentclass[amsmath,amssymb,showpacs,draft,preprint,floatfix]{revtex4}

\usepackage{graphicx}
\usepackage{latexsym}
\begin{document}
%\draft
\title{A number-phase Wigner function}
\author{H. Moya-Cessa}
\affiliation{INAOE, Coordinaci\'on de Optica, Apdo. Postal 51 y 216,
72000 Puebla, Pue., Mexico }
\date{\today}

\begin{abstract}

One of the most promininet quasiprobability functions in quantum mechanics
is the Wigner function that gives the right marginal probability functions
if integrated over position or momentum. Here we depart from the definition
of the position-momentum  Wigner function to, in analogy, construct a
number-phase Wigner function that, if summed over photon  numbers gives the
correct phase distribution and integrated over phase gives the right
photon distribution.
\end{abstract}

 \maketitle

Over the years, quantum phase space distributions have not only been useful
tools that allow to trancribe operator equations into c-number partial
differential equations, but have proved to have many other uses, for
instance in the reconstruction of the quantum state of a cavity field \cite{dav} or
the vibrational motion of ions \cite{Leibfried}. Among the most important ones, one can
mention the Wigner distribution function \cite{Wigner}, the Q function
\cite{Kano} and the  Glauber-Sudarshan P-function \cite{Sudarshan},
which belong to a more general one-parameter family of quantum phase space distributions.

Because they have an importance on their own, they have been extensively studied,
among others by Cahill and Glauber \cite{Cahill} and Agarwal and Wolf \cite{Agarwal}
and more recently by more authors in a theoretical \cite{Moya} and
experimental manner \cite{Leibfried}.

Here we would like to show that it can be constructed a Wigner distribution
function for phase and number in an analogous way in which is defined for
the position-momentum variables. We should point out that phase-number
Wigner functions have
been introduced by Vaccaro and Pegg \cite{Vaccaro} using the Pegg-Barnett \cite{Pegg}
phase formalism;
Luks and Perinova \cite{Luks} in an enlarged Hilbert space
and Vaccaro \cite{Vaccaro2}
by using the properties of the position-momentum Wigner function and applying them
to the phase-number Wigner function \cite{Vaccaro}. It should be mentioned that, althouhg
in the latest the marginal probabilities for phase and number are correct, figures in phase
space are not completely understood (for instance, a coherent state with phase
$\phi=0$ contains also a small "wave" at $\phi=\pi$ \cite{Vaccaro2}).

We start by writing the Wigner function for position and momentum in the form

\begin{equation}
W(\alpha) = \int C(\beta) \exp(\alpha\beta^*-\alpha^*\beta) d^2\beta,
\label{1}
\end{equation}
where the characteristic function, $C(\beta)$ is defined as

\begin{equation}
C(\beta) = Tr[\hat{D}(\beta)\hat{\rho}],
\label{2}
\end{equation}
with $\hat{\rho}$ the system's density matrix,
$\hat{D}(\beta)=\exp(\beta \hat{a}^\dagger-\beta^* \hat{a})$ the Glauber's
displacement operator, and $\hat{a}$ and $\hat{a}^\dagger$ the annihilation and creation
operators of the quantised field, respectively. One can further write it in terms of the
position and momentum operators such that the characteristic function times the Kernel
reads

\begin{equation}
\tilde{C}(\beta_x,\beta_p) = e^{-i\sqrt{2}(\beta_x\hat{p}-\beta_p\hat{x})}
e^{\alpha\beta^*-\alpha^*\beta},
\label{3}
\end{equation}
where we have set the frequency of the quantised electromagnetic field
and $\hbar$ equal to one.

In analogy to equation (\ref {3}) we can define the function
\begin{equation}
\tilde{C}_{\hat{n}-\hat{\Phi}}(k,\theta) = \frac{1}{2}Tr\left[\left(
\hat{D}_{\hat{n}-\hat{\Phi}}(k,\theta)e^{-i(k\phi-n\theta)}
+ c.c. \right)\hat{\rho}\right],
\label{4}
\end{equation}
where \cite{oneway}
\begin{equation}
\hat{D}_{\hat{n}-\hat{\Phi}}(k,\theta) = e^{i\theta k}e^{-i\theta\hat{n}}(\hat{V}^\dagger)^k ,
\label{5}
\end{equation}
with $\hat{V}^{\dagger}=\sum_{k=0}^{\infty}|k+1\rangle \langle k|$ the Susskind-Glogower operator
\cite{Susskind}. Because there is not a well defined phase operator, one can not use an expresion
of the form $\exp{[i(k\hat{\Phi}-\phi\hat{n})}]$, and we use instead a "factorized"
form in Eq. (\ref{5}).
Note that in order to produce a real Wigner function we added the complex conjugate in (\ref{4})
(because $n$ can not be a negative integer).
Eq. (\ref{1}) does not have this problem because the integrations over $\beta_x$ and $\beta_p$
are from $-\infty$ to $\infty$.  By writing the density matrix in the number
state basis,
\begin{equation}
\hat{\rho}= \sum^{\infty}_{m=0} \sum^{\infty}_{l=0} Q_{m,l}|m\rangle\langle l|,
\end{equation}
we obtain
\begin{equation}
\tilde{C}_{\hat{n}-\hat{\Phi}}(k,\theta) =
\frac{e^{i\theta k}}{2}\sum^{\infty}_{m=0} \sum^{\infty}_{l=0} Q_{m,l}
Tr[(\hat{V}^\dagger)^k e^{-i\theta\hat{n}}|m\rangle\langle l|]e^{-i(k\phi-n\theta)}+ c.c. .
\label{6}
\end{equation}
The double integration over the whole phase space in (\ref{1}) becomes here a sum and a single
integration
\begin{equation}
W(n,\phi)  =\frac{1}{(2\pi)^2} \sum^{\infty}_{k=-n} \int^{2\pi}_0
\tilde{C}_{\hat{n}-\hat{\Phi}}(k,\theta)d\theta. \label{7}
\end{equation}
Inserting equation (\ref{6}) into (\ref{7}) we obtain
\begin{equation}
W(n,\phi) =\frac{1}{4\pi}
\sum^{\infty}_{k=-n} \left(Q_{n,n+k}e^{-ik\phi}+Q_{n+k,n}e^{ik\phi} \right).
\label{8}
\end{equation}
It is easy to show that integrating (\ref{8}) over the phase $\phi$
\begin{equation}
\int^{2\pi}_0 W(n,\phi)  d\phi = Q_{n,n}=P(n),
\label{9}
\end{equation}
gives the photon distribution. And adding (\ref{8}) over $n$
\begin{equation}
\sum^\infty_{n=0} W(n,\phi)  =\frac{1}{4\pi} \sum^\infty_{n=0}
\sum^{\infty}_{k=-n}\left( Q_{n,n+k}e^{-ik\phi} +Q_{n+k,n}e^{ik\phi} \right).
\label{10}
\end{equation}
that may be rewritten as (by doing $m=k+n$)
\begin{equation}
\sum^\infty_{n=0} W(n,\phi)  =\frac{1}{2\pi} \sum_{n=0}^\infty
\sum_{m=0}^\infty Q_{n,m} e^{-i(m-n)\phi}
\label{11}
\end{equation}
produces the  correct phase distribution.
It is worth to note that for a number state $|M\rangle$ equation (\ref{8}) reduces to
$W(n,\phi) =  \delta_{nM}/2\pi$, i.e. it is different from zero only for $n=M$ as
it should be expected.
\subsection{Coherent state}
The phase-number Wigner function for a coherent state
\begin{equation}
|\alpha\rangle = e^{-|\alpha|^2/2}\sum_{m=0}^{\infty}\frac{\alpha^m}{\sqrt{m!}}|m\rangle
\label{coh}
\end{equation}
is given by
\begin{equation}
W(n,\phi) = \frac{e^{-|\alpha|^2}\alpha^{n}}{2\pi\sqrt{n!}}\sum_{k=0}^{\infty}
\frac{\alpha^k \cos[(n-k)\phi]}{\sqrt{k!}}.
\end{equation}
In Fig. (\ref{fig1}) it is plotted the phase-number Wigner function for an amplitude
$\alpha=4$ and $\phi=0.5$. Besides being always positive, it may be noted a smooth behavior.
\subsection{Phase-number Wigner function for a Schr\"odinger cat}
A Schr\"odinger cat, or a superposition of two coherent states may be given by the state
\begin{equation}
|\psi_\alpha \rangle = \frac{1}{N_\alpha} (|\alpha \rangle + |-\alpha\rangle),
\label{cat}
\end{equation}
with $N_{\alpha}$ is the normalization constant. By writing $|\psi_{\alpha}\rangle$
in a number states basis, one finds the phase-number Wigner function as
\begin{equation}
W(n,\phi) = \frac{e^{-|\alpha|^2}\alpha^{n}(1+(-1)^n)}{2\pi\sqrt{n!}N_{\alpha}}
\sum_{k=0}^{\infty}\frac{\alpha^k (1+(-1)^k) \cos[(n-k)\phi]}{\sqrt{k!}}.
\label{wig}
\end{equation}
In Fig. (\ref{fig2}) Eq. (\ref{cat}) is plotted.
The oscillatory behavior in the photon number and
the negativity , both proper of a Schr\"odinger cat may be clearly observed.
Also the two peaks of the coherent states.
\subsection{A special superposition of number states}
Let us consider the state
\begin{equation}
|\phi_M\rangle = \frac{1}{\sqrt{M+1}}\sum_{m=0}^M e^{i m \phi_0}|m\rangle.
\label{phi}
\end{equation}
This state tends to have a completely well defined phase as $M$ tends to infinity.
For this state the phase-number Wigner function reads
\begin{equation}
W(n,\phi) = \frac{1}{2(M+1)\pi}\sum_{k=0}^M\cos[(n-k)(\phi-\phi_0)],
\end{equation}
that may be put in the form \cite{Grad}
\begin{equation}
W(n,\phi) = \frac{1}{2(M+1)\pi}\cos\left[\left(\frac{M}{2}-n\right)(\phi-\phi_0)\right]
\sin\left[\frac{M+1}{2}(\phi-\phi_0)\right]
\csc\left(\frac{\phi-\phi_0}{2}\right)
\label{wphi}
\end{equation}
In Fig. (\ref{fig3}) it is plotted (\ref{wphi}) for $M=20$ and $\phi_0=0.7$.
It shows a well defined phase. It is also seen that as $\phi$ approaches the value
$\phi_0 $ the maximum value for the phase-number Wigner function for the state (\ref{phi})
is obtained. From (\ref{wphi}) it may be shown that this value is $1/2\pi$.
By adding over $n$ equation (\ref{wphi}) the phase distribution is obtained
\begin{equation}
P(\phi)=\frac{1}{2(M+1)\pi}
\sin^2\left[\frac{M+1}{2}(\phi-\phi_0)\right]
\csc^2\left(\frac{\phi-\phi_0}{2}\right)
\end{equation}
that corresponds to the phase distribution for the state (\ref{phi}).

In conclusion we have constructed a phase-number Wigner function that gives the correct
marginal probabilities if added over the photon number or integrated over phase.
It was constructed as an analogy to the definition of the position-momentum Wigner function
given in (\ref{1}).

This work was partially supported by CONACYT (Consejo Nacional de
Ciencia y Tecnolog\'\i a, Mexico).

\begin{figure}[hbt]
\caption{\label{fig1}
Phase-number Wigner distribution function for the state (\ref{coh}).
The amplitude of the coherent state is $|\alpha|=4$ and the phase is $0.5$}

\caption{\label{fig2}
Phase-number Wigner distribution function for the Schr\"odinger cat state (\ref{cat}).
$\alpha=4$.}
\caption{\label{fig3}
Phase-number Wigner function for the quasi-phase state (\ref{phi}) for $M=20$ and
$\phi_0=0.7$. }

\end{figure}


\begin{thebibliography}{aaaa}

\bibitem{dav} L.Lutterbach and L. Davidovich, Phys. Rev. Lett. {\bf 78}, 2547 (1997).
\bibitem{Leibfried}
D. Leibfried, D.M. Meekhof, B.E. King, C. Monroe, W.M. Itano
and D.J. Wineland, Phys. Rev. Lett. {\bf 77}, 4281 (1996).
\bibitem{Wigner}
E.P. Wigner, Phys. Rev. {\bf 40}, 749 (1932).
\bibitem{Kano}
Y. Kano, J. Phys. Soc. Japan {\bf 19}, 1555 (1964).
\bibitem{Sudarshan}
R.J. Glauber, Phys. Rev. Lett {\bf 10}, 84 (1963); E.C.G. Sudarshan.
Phys. Rev. Lett. {\bf 10}, 277 (1963).
\bibitem{Cahill}
K.E. Cahill and R.J. Glauber, Phys. Rev. {\bf 177}, 1857 (1969).
\bibitem{Agarwal}
G.S. Agarwal and E. Wolf, Phys. Rev. D{\bf 2}, 2161 (1970).
\bibitem{Moya}
H. Moya-Cessa and P.L. Knight, Phys. Rev. A{\bf 48}, 2479 (1993).
\bibitem{Vaccaro}
J.A. Vaccaro and D.T. Pegg, Phys. Rev. A {\bf 41}, 5156 (1990).
\bibitem{Pegg} D.T.  Pegg and S.M. Barnett, Europhys. Lett. {\bf 6}, 483 (1988);
S.M. Barnett and D.T. Pegg, J. Mod. Optics {\bf 36}, 7 (1989); D.T. Pegg and S.M. Barnett,
Phys. Rev. A {\bf 39}, 1665 (1989).
\bibitem{Luks}
A. Luks and V. Perinova, Phys. Scr. {\bf T48}, 94 (1993).
\bibitem{Vaccaro2} J.A. Vaccaro, Opt. Commun. {\bf 113} 421 (1995);
 Phys. Rev. A {\bf 52}, 3474 1995.
\bibitem{oneway}
Note that $\hat{D}^\dagger_{\hat{n}-\hat{\Phi}}(k,\theta)\hat{D}_{\hat{n}-\hat{\Phi}}(k,\theta)$
is equal to one,
but $\hat{D}_{\hat{n}-\hat{\Phi}}(k,\theta) \hat{D}^\dagger_{\hat{n}-\hat{\Phi}}(k,\theta)$ is not.
\bibitem{Susskind}
L. Susskind and J. Glogower, Physics {\bf 1}, 49 (1964).
\bibitem{Grad} I.S. Gradshteyn and I.M. Ryzhik, {\it Table of integrals, series, and products}
Academic Press, Inc., London 1980), p. 29.
%\bibitem{note1}  One can of course find a solution if $\Omega $ is allowed
%to take on complex values, which is equivalent to choosing a nonzero laser
%phase $\theta _{0}$ in eq. (\ref{ion1}). In this case, we have four parameters
%($\Omega ,\delta ,\eta ,\theta _{0}$), only three of which can be chosen
%independently.

\end{thebibliography}
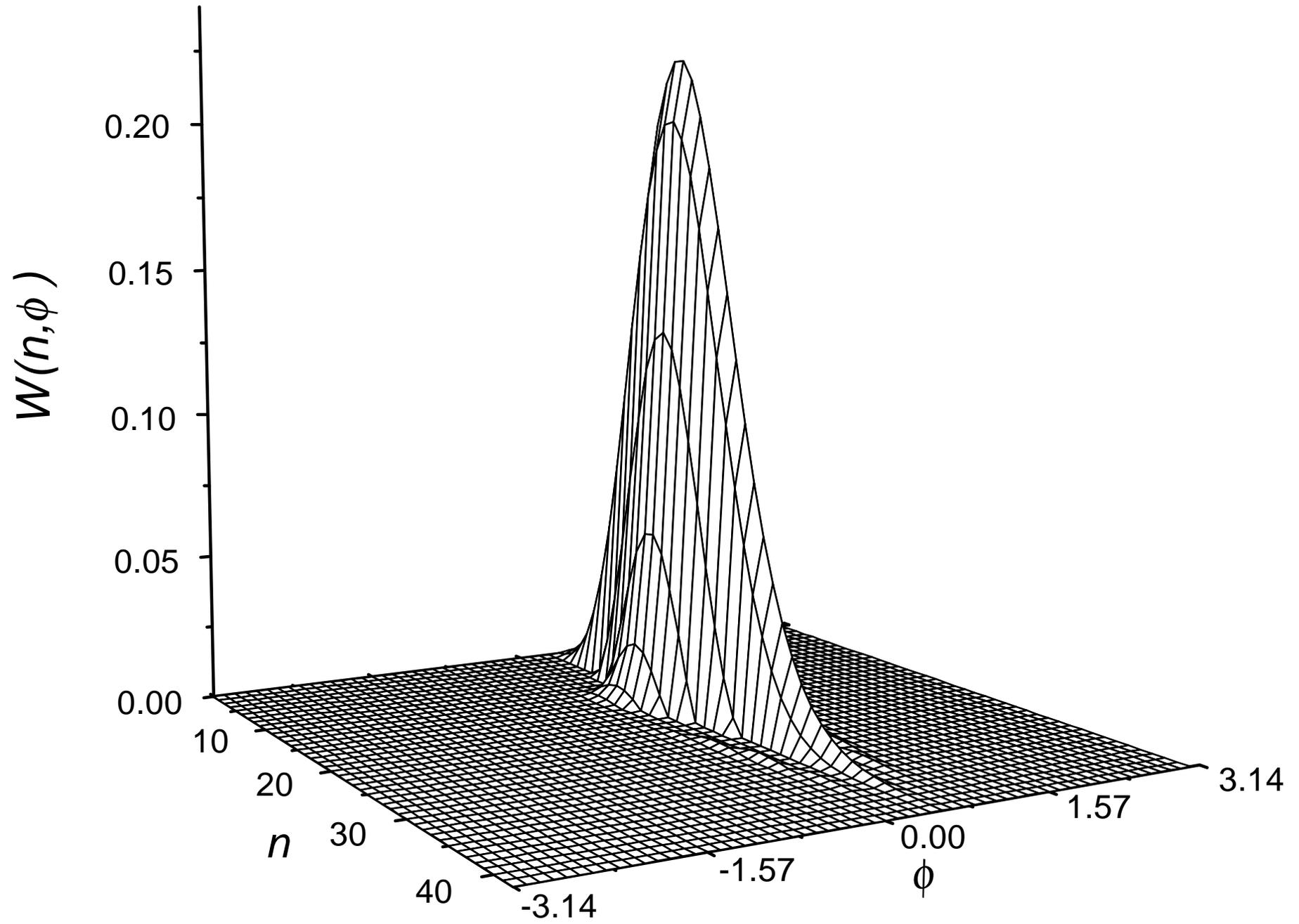
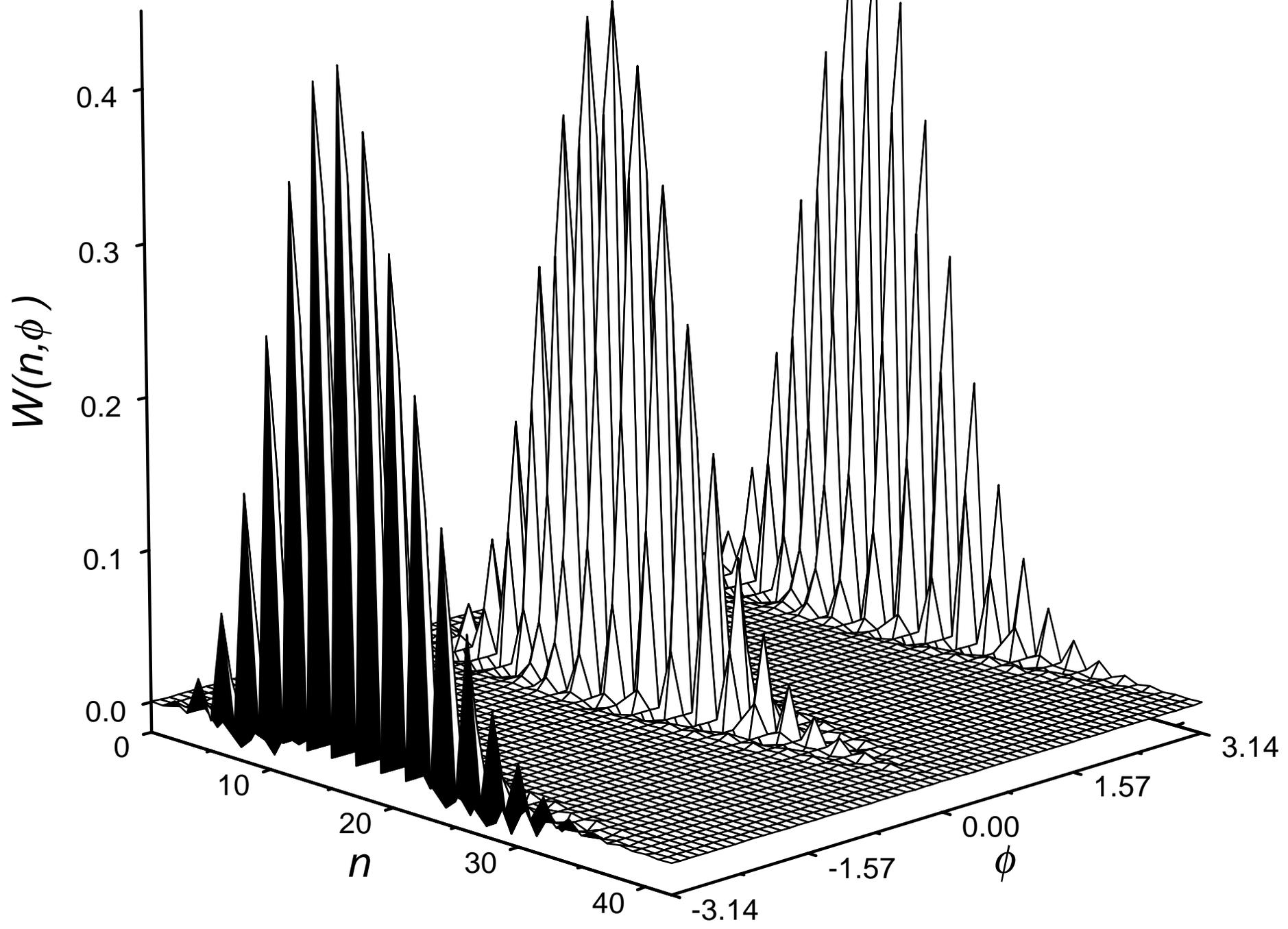
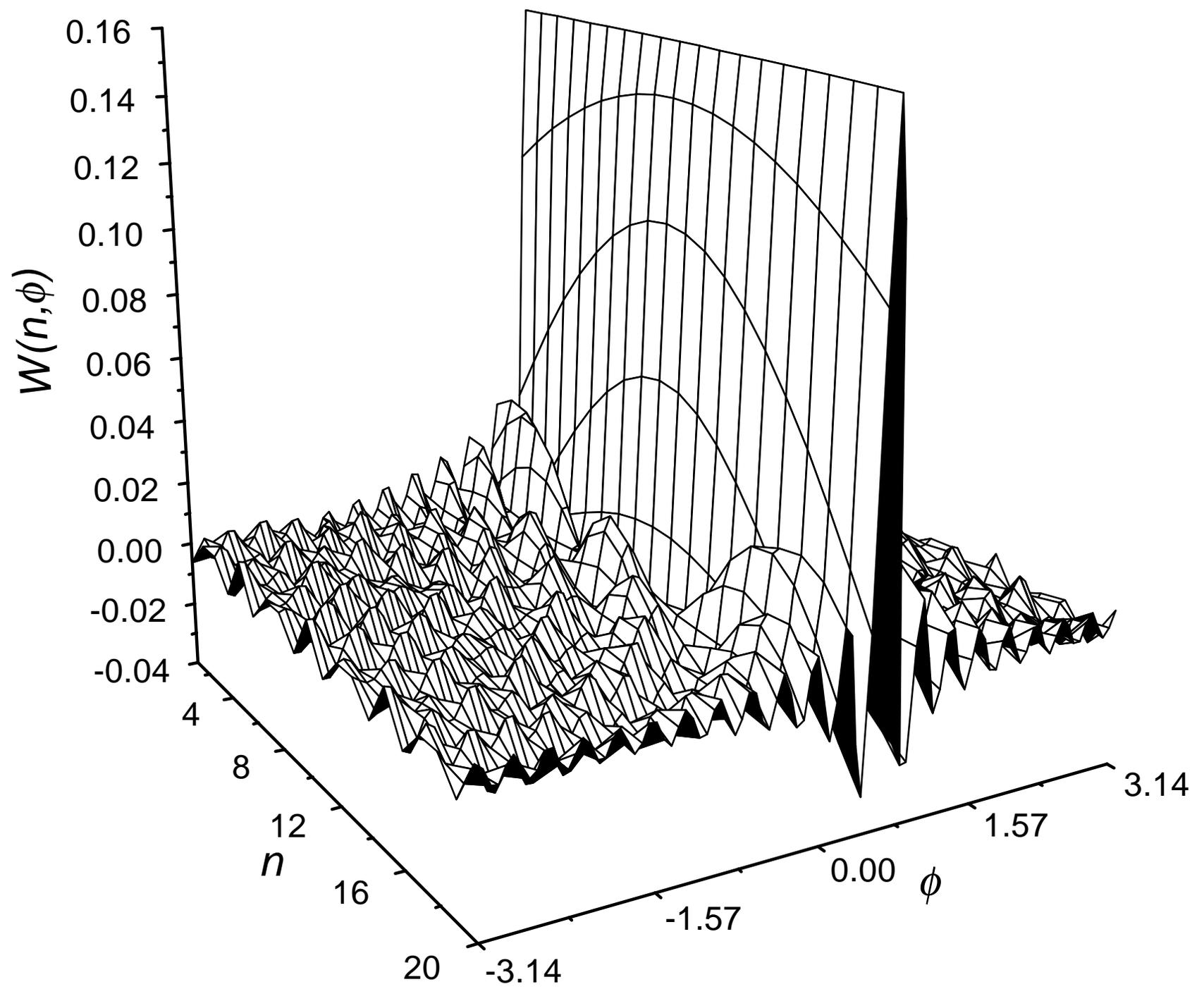
\end{document}